\documentclass[12pt, amsfonts, amssymb]{article}

\usepackage{amssymb}
\usepackage{amsmath}
\usepackage{latexsym}

\newcommand{\CC}{{\mathbb C}}
\newcommand{\RR}{{\mathbb R}}
\newcommand{\ZZ}{{\mathbb Z}}
\newcommand{\PP}{{\mathbb P}}

\newcommand{\ra}{\rightarrow}

\newcommand{\eps}{\epsilon}

\newcommand{\bpsi}{{\bar\psi}}
\newcommand{\cD}{{\mathcal D}}
\newcommand{\cL}{{\mathcal L}}

\title{\bf \large Topological Disorder Operators in Three-Dimensional
Conformal Field Theory}
\author{Vadim Borokhov\thanks{borokhov@theory.caltech.edu}, Anton Kapustin\thanks{kapustin@theory.caltech.edu}, Xinkai Wu\thanks{xinkaiwu@theory.caltech.edu}\\
\it California Institute of Technology, Pasadena, CA 91125, USA
}

\begin{document}
\begin{titlepage}

\renewcommand{\thepage}{ }

\maketitle

\begin{abstract}

Many abelian gauge theories in three dimensions flow to interacting
conformal field theories in the infrared. We define a new class
of local operators in these conformal field theories which are not
polynomial in the fundamental fields and create
topological disorder. They can be regarded as higher-dimensional
analogues of twist and winding-state operators in free 2d CFTs. 
We call them monopole 
operators for reasons explained in the
text. The importance of monopole operators is that in the Higgs phase, 
they create Abrikosov-Nielsen-Olesen vortices. 
We study properties of these operators in three-dimensional QED using 
large $N_f$ expansion. In particular, we show that monopole operators
belong to representations of the conformal group whose primaries have
dimension of order $N_f$. We also show that monopole operators transform
non-trivially under the flavor symmetry group, with the precise representation
depending on the value of the Chern-Simons coupling.

\end{abstract}

\vspace{-6.5in}

\parbox{\linewidth}
{\small\hfill CALT-68-2391}

\end{titlepage}


\section{Introduction}

One of the most fascinating problems in quantum field theory is understanding
non-perturbative equivalences (``dualities'') between superficially
very different theories. A classic example is the quantum equivalence of the massive Thirring and sine-Gordon models~\cite{Coleman,Mandelstam}.
More recently, a number of dualities has been conjectured for supersymmetric
gauge theories in three and four dimensions. The earliest proposal of
this sort is the S-duality of $N=4$ $d=4$ super-Yang-Mills 
theory~\cite{MO,WO,Osborn}.
A decade and a half later, N.~Seiberg proposed a dual description for the 
4d CFT which arises as the infrared limit of $N=1$ $d=4$
super-QCD~\cite{Seiberg}. The dual theory is again the infrared limit of 
an $N=1$ $d=4$ gauge theory. 
This proposal generated tremendous excitement, and soon many other candidate dualities have been found (see Refs.~\cite{Peskin,CCM} for a review). 
Later it was realized that 
many field-theoretic dualities follow from string theory dualities.

Until now, all dualities in dimensions
higher than two remain conjectural, and the physical reason for their existence is not completely understood. On the other hand, 2d dualities have a rather
transparent physical meaning. For example, the sine-Gordon model has
topological solitons (kinks), and it can be shown that a certain local
operator which creates a kink satisfies the equations of motion of the
massive Thirring model~\cite{Mandelstam}. 
It is believed that many higher-dimensional
dualities arise in a similar manner, by ``rewriting'' the theory in terms
of operators which create topological disorder. But it proved very hard
to make this idea precise. 

There are several related difficulties that one encounters in dimension
higher than two. First of all, interesting higher-dimensional
dualities involve gauge theories. This implies that in order to write
down an operator describing the dual degrees of freedom,
one has to work in an enlarged state space which includes the
unphysical degrees of freedom of both the original and the dual gauge fields. It is not known how to construct such an enlarged space.
Fortunately,
there are non-trivial examples of dualities in three dimensions 
(so called 3d mirror pairs~\cite{IS}) for some of which the dual theory has a 
trivial gauge group. In this case one can hope to construct the
operators describing the dual degrees of freedom directly in the 
state space of the original gauge theory. 

The second difficulty is that it is hard to construct
topological disorder operators in interacting fields theories. For
example, it is believed that 3d mirror symmetry arises when one
rewrites three-dimensional supersymmetric QED in terms of local operators 
which create Abrikosov-Nielsen-Olesen vortices~\cite{five}. 
This means that such operators are monopoles.
However, it was never clear how to define monopole operators in
SUSY QED. A proposal in this direction was made by M.~J.~Strassler and 
one of the authors~\cite{KS}, but it was only partially
successful.

In this paper we will address the second issue in a simple toy 
model: three-dimensional QED with $N_f$ flavors of fermions. 
This theory is believed to flow to an interacting conformal
fixed point for large enough $N_f$ (this is discussed in more detail in the
next section). The theory is not supersymmetric and is not expected to
possess a simple dual. Nevertheless, we believe it is a useful
exercise to define monopole operators in this simple model
and learn to work with them. Besides, monopole operators
are rather interesting beasts even in the non-supersymmetric
case. First of all, these are the first examples of local operators
in a three-dimensional CFT which are not polynomial in the fundamental
fields. Thus our construction can be regarded as a generalization of the
vertex operator construction from free 2d CFT to an interacting
3d CFT. Second, we show that because of fermionic zero modes
monopole operators transform in a non-trivial
representation of the flavor group whose size depends on the Chern-Simons
coupling. Monopole operators in supersymmetric QED and their role
in mirror symmetry will be discussed in a forthcoming publication.

\section{Review of three-dimensional QED}

The action of three-dimensional QED in the Euclidean space is given by
$$
L_{QED}=\int d^3 x \left(\frac{1}{4e^2} F_{\mu\nu} F^{\mu\nu} +
\psi^\dagger_j \left(\sigma \cdot iD_A\right) \psi^j\right).
$$
Here $A$ is the $U(1)$ gauge field, $F=dA$ is the field-strength 2-form,
$D_A$ is the corresponding covariant derivative,
and $\psi^j$ is a complex two-component spinor. The index $j$ runs from 
$1$ to $N_f$. 

In three dimensions one can add to the action a Chern-Simons term
$$
L_{CS}=\frac{ik}{4\pi} \int d^3 x\ \eps^{\mu\nu\rho} A_\mu 
\partial_\nu A_\rho.
$$
Such a term breaks parity invariance of the theory.
We will assume that the gauge group is compact (i.e.
$U(1)$ rather than $\RR$). Naively, this requires the Chern-Simons
coupling $k$ to be an integer, to avoid global anomalies. The real
story is slightly more complicated. When $N_f$ is odd, the fermionic path
integral is anomalous. The anomaly is the same as the anomaly due
to a Chern-Simons term with $k=1/2$. Thus cancellation of global
anomalies requires
$$
k-\frac{N_f}{2}\in \ZZ.
$$
In particular, for odd $N_f$ the Chern-Simons coupling must be non-zero,
and parity is broken. This is known as parity anomaly~\cite{Redlich}.

The gauge coupling $e$ has dimension $m^{1/2}$. Thus the theory is
super-renormalizable and free in the ultraviolet. In fact, its UV
behavior is so good that no renormalization of the Lagrangian is required.
Contrariwise, 3d QED is strongly coupled in the infrared (perturbative
expansion is really an expansion in powers of $e^2/p$, where $p$ is
the momentum scale).
It is natural to assume that the
low-energy limit of this theory is a non-trivial CFT, but this has not
been conclusively demonstrated. However, the statement holds
to all orders in $1/N_f$ expansion~\cite{AP,JT,Templeton,AH}. 
In fact, in the limit 
$N_f\ra\infty$ the infrared theory becomes weakly coupled, and
conformal dimensions of all fields can be computed order by order 
in $1/N_f$. For example, the IR dimension of $\psi^j$ is 
canonical (i.e. the same as the UV dimension) up to corrections of 
order $1/N_f$. 

More interestingly,
the IR dimension of $F_{\mu\nu}$ is $2$ to all orders in $1/N_f$.
To understand why this is the case, consider a current
$$
J^\mu=\eps^{\mu\nu\rho} F_{\nu\rho}.
$$
It is identically conserved by virtue of the Bianchi identity.
A priori, this current could either be a primary field, or a descendant
of the primary field. In the UV, the latter possibility is realized,
since we can write 
\begin{equation}\label{dualph}
J^\mu=\partial^\mu \sigma,
\end{equation}
where $\sigma$ is a free scalar field. The scalar $\sigma$ is usually
referred to as the dual photon. It has dimension $1/2$ (as befits a free scalar in three dimensions), while $J^\mu$ and $F_{\mu\nu}$ have dimension $3/2$. 
On the
other hand, in the IR an equation like Eq.~(\ref{dualph}) cannot hold.
Indeed, Eq.~(\ref{dualph}) implies that $F_{\mu\nu}$ obeys the free Maxwell equation, which clashes with the assumption that there are massless charged particles in the infrared. (We assume here that the fermions do not get a mass due to some non-perturbative effect, see a discussion below.) 
This strongly suggests that in the IR limit $J^\mu$ is a primary field. 
It is well known that 
in a unitary 3d CFT a conserved primary current has dimension $2$. 
Hence the IR dimension of $J^\mu$ and $F_{\mu\nu}$ is $2.$ 
This conclusion can also
be reached by directly studying the perturbative expansion in powers 
of $1/N_f$~\cite{AH}.

Note that the difference between the UV and IR dimensions of $F$
is of order $1$, and therefore the IR fixed point is far from the UV
fixed point, even in the limit $N_f\ra\infty$. In this respect,
the situation is very different from the Banks-Zaks-type theories
in four-dimensions~\cite{BZ}, where the IR dimensions of {\it all} operators are very close to their UV dimensions.

The physics of 3d QED at finite $N_f$ remains controversial. The conventional
approach is to study the system of Schwinger-Dyson equations truncated 
in some way and look for symmetry-breaking solutions. For simplicity,
let us focus on the case of zero Chern-Simons coupling and even $N_f$.
It has been claimed that at finite $N_f$ flavor symmetry and parity are spontaneously broken by a dynamical mass for the fermions, and the infrared limit is a free theory of photons~\cite{Pisarski}. The majority of such studies indicate that this happens for $N_f$ smaller than a certain critical value of order $6$ or $7$~(see e.g.~\cite{ANW,Nash,Maris}).
There are also claims that dynamical mass 
generation takes place for all $N_f$ but is exponentially small for large
$N_f$ and therefore invisible in $1/N_f$ expansion~\cite{Pisarski,PW,CPW}.
It must be stressed that the results of such studies depend on the way one truncates an infinite system of Schwinger-Dyson equations, a procedure which 
cannot be fully justified. Lattice simulations of 3d QED have been
inconclusive so far. 

In this paper we will be
interested in the large $N_f$ limit, and therefore the behavior at
finite $N_f$ will be unimportant. Note also that in the $N=2$ and $N=4$
supersymmetric cases the situation is better, in the sense that one can argue
the existence of a non-trivial CFT at the origin of the quantum moduli
space for all $N_f$. 

\section{Defining monopole operators}

\subsection{A preliminary definition}

As mentioned above, three-dimensional QED possesses an interesting conserved current, the dual of the field strength:
$$
J^\mu=\frac{1}{4\pi}\eps^{\mu\nu\rho} F_{\nu\rho}.
$$
Its conservation is equivalent to the Bianchi identity $dF=0.$
The corresponding charge is called the vortex charge, because in the
Higgs phase it is carried by the Abrikosov-Nielsen-Olesen (ANO)
vortices. The vortex charge is integral if the gauge field $A$ is
a well-defined connection on a $U(1)$ principal bundle. Loosely speaking,
we would like to construct a vortex-creating operator. 
But in an interacting conformal field theory, 
it does not make sense to say that an 
operator is creating a particle. A vortex-creating operator will 
be defined as an operator with a unit vortex charge. This means that the
OPE of such an operator with $J^\mu$ has the form
$$
J^\mu(x) O(0)\sim \frac{1}{4\pi}\frac{x^\mu}{|x|^3} O(0)+
{\rm less\ singular\ terms}.
$$
Such operators can be organized in the representations of the conformal
group. In a unitary theory local operators must transform
according to lowest-weight representations, i.e.
those representations in which the dimension of operators is bounded
from below. The operator with the lowest dimension is called a conformal
primary. It is standard to label a representation by the spin and dimension 
of its primary. Our problem can be formulated
as follows: determine the spin, dimension, and other quantum numbers
of primaries with a given vortex charge.

In the path integral language, an insertion of an operator with vortex
charge $n$ at a point $p$ is equivalent to integrating
over gauge fields which have a singularity at $x=p$ such that
the magnetic flux through a 2-sphere surrounding $x=p$ is $n$.
To be consistent, one must regard charged matter fields as sections
of a non-trivial line bundle on the punctured $\RR^3.$ Thus
an insertion of a vortex-creating operator causes a change in the topology
of fields near the insertion point. In what follows we will use the
terms ``vortex-creating operator'' and ``monopole operator'' 
interchangeably.

This way of defining topological
disorder operators is familiar from 2d CFT. For example, a twist
operator for a free fermion in 2d is defined by the condition that the
fermion field changes sign as one goes around the insertion 
point~\cite{Ginsparg}.
Another example is afforded by the theory of a periodic free boson in 2d.
This theory has winding states, and the corresponding operators
create a logarithmic singularity for the boson field.  
Thus our monopole operators can be regarded as three-dimensional analogues
of twist operators or winding-state operators.

In the two-dimensional case one can loosely say that a winding-state
operator creates a kink. The precise meaning of this statement is the
following. Consider a perturbation of the free boson theory by
a periodic potential, say, a sine-Gordon potential. The resulting
massive theory has multiple vacua and topological excitations (kinks) interpolating between neighboring vacua. The operator which carries
winding number one has non-zero matrix elements between the vacuum
and the one-kink state.

Similarly, one can loosely say that a monopole operator creates
an ANO vortex. To make this statement precise, one
has to go to the Higgs phase (for example, by adding charged scalars
with an appropriate potential). In the Higgs phase, the magnetic
flux emanating from the insertion point of the monopole operator
is squeezed into a thin tube. This tube is the world-line of a vortex.

\subsection{A more precise definition}

The definition of monopole operators given above is not yet complete.
In effect, we have defined an insertion of a monopole operator by 
requiring that the gauge field strength have a particular singularity
at the insertion point. However, we did not specify the behavior
of the matter fields near the insertion point. In fact, we expect
that there are many operators which carry the same vortex charge, and
they differ precisely by the behavior of fields at the insertion point. 
Another difficulty is that the IR theory is strongly coupled, and it seems
hard to compute correlators involving monopole operators.

The first difficulty can be circumvented using radial quantization. 
It is a general feature of CFT in any dimension that local operators
are in one-to-one correspondence with states in the Hilbert space of
the radially quantized theory. This follows from the fact that one
can use a conformal transformation to map an insertion point of an
operator to infinity. In this way one trades a local operator for
an incoming or out-going state. In the case of monopole operators,
such a mapping takes an operator with vortex charge $n$ to a state on 
${\bf S}^2\times \RR$ with a magnetic flux $n$ through ${\bf S}^2$.
Here $\RR$ is regarded as the time direction.
Classifying states of a CFT on ${\bf S}^2\times \RR$ with a given
vortex charge is certainly a well-defined problem. Furthermore,
the radially quantized picture is the most convenient one for
computing correlators which involve two monopole operators with
opposite vortex charges and an arbitrary number of ordinary operators. 
By mapping the insertion of
a monopole operator to an in-going state and the insertion of an anti-monopole
operator to an out-going state, one reduces the problem
to computing a particular matrix element of a product of several ordinary
operators. A particularly important
special case is the three-point function which
involves a monopole operator, an anti-monopole operator, and a 
conserved current. Knowledge of such correlators allows one
to read off the quantum numbers of a monopole operator. For example,
in order to determine the dimension of an operator, one has to
compute the expectation value of the stress-energy tensor in the
corresponding state. This approach is familiar from 2d CFT,
where it is used to compute the quantum numbers of twist 
operators (see e.g. Ref.~\cite{Ginsparg}). 

Of course, if one desires to compute four-point functions of monopole
operators, mapping two of the insertion points to infinity does not help 
very much. In the case of 2d CFT, one has to use tricks special to the theory in question in order to compute four-point functions of topological 
disorder operators. In this paper, we will be content with studying the OPE of monopole operators with conserved currents, and leave the study of four-point 
functions to future work.

The second difficulty can be avoided by working in the large $N_f$ limit.
It is a general feature of this limit that the gauge field does not
fluctuate, and can be treated classically~\cite{AP,JT,AH}. 
This can be seen as follows.
The infrared limit in 3d QED is simply the limit $e\ra\infty$. This
is literally true, because no renormalization of the Lagrangian is
required. Thus one can simply drop the kinetic term for the gauge field.
Integrating out the fermions then gives an effective action for the
gauge field of order $N_f$. For example, when expanded around a trivial
background, this action looks like
$$
N_f \int \left(F_{\mu\nu}\,\Box^{-1/2} F^{\mu\nu} + 
{\rm higher\!-\! order\ terms}\right) d^3 x.
$$
Thus the effective Planck constant is of order $1/N_f$, and in the
large $N_f$ limit the size of gauge-field fluctuations is order $1/N_f$.
Moreover, if we absorb a factor of $N_f^{1/2}$ into $F,$ we see
that self-interactions of $F$ are suppressed in the large $N_f$ limit. 
In other words, $N_f^{1/2} F$ is a Gaussian field in the large $N_f$ limit.
It is this
line of reasoning that allows one to show that the infrared CFT is weakly
coupled in the large $N_f$ limit. The argument also applies
to CFT on ${\bf S}^2\times \RR$ with a flux. Thus we can regard the
gauge field as a classical background. 

It is very plausible that the
saddle point of the effective action for $F$ on ${\bf S}^2\times \RR$
is rotationally symmetric. Therefore we can assume that the classical
background is simply a constant magnetic flux on ${\bf S}^2$.

The above discussion reduced our problem to studying free fermions
on ${\bf S}^2\times \RR$ in the presence of a constant magnetic flux.
This is almost a textbook problem, and everything of interest can be
computed. For example, finding the dimension of a monopole operator 
is equivalent to computing the Casimir energy of free fermions on ${\bf S}^2$
with a flux. It is a priori clear that this energy scales like $N_f.$
There are corrections to this result, which can be computed
by taking into account the fluctuations of the gauge field. However, such
effects are suppressed by powers of $1/N_f.$

The above discussion contains a gap as regards gauge invariance of
monopole operators. Gauge-invariance of a local operator is equivalent
to gauge-invariance of the corresponding state in the
radially quantized picture.
In other words, the state must satisfy the Gauss' law. Gauss' law
in QED on ${\bf S}^2\times \RR$ comes from varying the action with
respect to the ``time-like'' component of the gauge field $A$. In the
limit $e\ra\infty$ it simply reads
$$
\rho(x) \vert \Phi\rangle =0,
$$
where 
$$\rho(x)=\sum_j\psi_j^\dag(x)\psi^j(x)
$$ 
is the electric charge density operator.
In particular, the total electric charge of a gauge-invariant state must
be zero. The latter is a standard fact about gauge theory on a compact space,
valid irrespective of the value of $e$. The definition of the
electric charge operator involves normal-ordering ambiguities,
which will be dealt with below. Note also that the inclusion of the
Chern-Simons term in the action modifies the Gauss' law constraint into
$$
\left(\rho + \frac{k}{4\pi}\eps^{ij} F_{ij}\right)\vert\Phi\rangle =0.
$$
In particular, the total electric charge of the matter modes must be
equal to $-k$ times the vortex charge. In this way (and only in this way)
the Chern-Simons term will affect the physics at large $N_f$.
 
\section{Properties of monopole operators}
 
\subsection{Radial quantization in the presence of a flux}

As explained in the previous section, at large $N_f$ all properties of
monopole operators can be deduced from the properties of free fermions on
${\bf S}^2\times \RR$ in a constant background magnetic flux. In this 
subsection we summarize the physics of this system, with detailed
derivations relegated to the Appendix.

The spectrum of the Dirac Hamiltonian on ${\bf S}^2\times \RR$ with
$n$ units of magnetic flux is given by
$$
E_p=\pm {\sqrt{p^2+p|n|}},\qquad p=0,1,2,\ldots .
$$
The degeneracy of the $p$-th eigenvalue is $2j_p+1$, where
$$
j_p=\frac{1}{2}(|n|-1)+p.
$$
These $2j_p+1$ states transform as an irreducible representation of
the rotation group $SU(2)_{rot}$.

The presence of $n$ states with zero energy is particularly important. 
The existence of at least $n$ zero modes is dictated by the Atiyah-Singer 
index theorem applied to the Dirac operator on ${\bf S}^2$ coupled to the
magnetic field. 

In the case of a unit magnetic flux ($|n|=1$), we have a single fermionic
zero mode with zero spin. Thus a spinor is converted into a scalar
due to the non-trivial topology of the magnetic monopole. This 
scalar-spinor transmutation is well known in other contexts;
in particular it plays an important role in the conjectured S-duality
of $N=4$ $d=4$ super-Yang-Mills. For general $n$, the fermionic zero
modes transform in an irreducible representation of $SU(2)_{rot}$ with spin
$j=(|n|-1)/2.$ We will discuss in detail the case when $n=\pm 1$, 
and then comment on the higher-$n$ case.

Let us denote the fermionic
annihilation operators by $c^i_{pm}$, where $i=1,\ldots,N_f$ is the
flavor index, $p=1,2,\ldots,$ labels the energy eigenspaces as above,
and $m=-j_p,-j_p+1,\ldots,j_p,$ labels the states within the $p$-th
energy eigenspace. The fermion annihilation operators corresponding to
$p=0$ will be denoted simply by $c^i_0.$
The Hilbert space of the theory is the tensor product of the Hilbert
space of zero modes and the Hilbert space of all other modes.
The latter is simply a fermionic Fock space with a unique vacuum
$|vac\rangle_+$ which satisfies
$$
c^i_{pm}|vac\rangle_+=0,\quad p>0,\forall i,m.
$$
This vacuum state is rotationally invariant.

The Hilbert space of zero modes is also a Fock space of dimension
$2^{N_f}$, with the vacuum vector which we denote $|vac\rangle_0$.
It is spanned by the vectors
$$
|vac\rangle_0,\ c^{i\dagger}_{0}|vac\rangle_0,\ 
c^{i_1\dagger}_0 c^{i_2\dagger}_0|vac\rangle_0,\ 
\ldots,\ 
c^{i_1\dagger}_0 c^{i_2\dagger}_0\ldots c^{i_{N_f}\dagger}_0
|vac\rangle_0.
$$
All these states are degenerate in energy, and none is 
a preferred vacuum. Since the zero modes have spin zero, all the
ground states are rotationally invariant. 
We conclude that the radially-quantized theory of free fermions 
has a $2^{N_f}$-fold degenerate ground state.

However, we still need to impose the Gauss' law. The charge density
operator receives contributions from both zero and non-zero modes.
The part due to non-zero modes can be defined using the obvious
normal-ordering prescription.
If we put all non-zero modes in the vacuum state, then the
charge density due to non-zero modes vanishes. It remains to analyze the
contribution from zero modes. Naively, it seems that the Fock
vacuum must be assigned zero electric charge. If this were the case,
then the states obtained by acting on the vacuum with zero mode creation operators would have positive charge, and therefore would not be
gauge-invariant. But because of normal-ordering ambiguities, 
the situation is more interesting.

As stressed above, the Fock vacuum for the zero
modes is not that special. The completely filled state appears to be
an equally good candidate for a state with vanishing electric charge. The
two just differ by a change in the normal ordering prescription.
A statement which is independent of the normal-ordering prescription
is that the electric charge of the filled state exceeds the charge of the
vacuum by $N_f$. If one wants to be ``democratic'', one has 
to assign charge $-\frac{1}{2}N_f$ to the vacuum and charge $\frac{1}{2}N_f$
to the filled state. A similar symmetric charge assignment has been
advocated by Jackiw and Rebbi in their pioneering study of fermions bound 
to solitons, on the grounds on charge-conjugation symmetry~\cite{JR}.

The precise argument for the symmetric charge assignment goes as follows.
Charge conjugation maps a monopole to an anti-monopole and by itself does
not tell us anything. But CP transformation maps a monopole to itself.
If we want to quantize in a CP-invariant
manner, we must assign opposite electric charges to states related by CP.
Since CP takes annihilation operators into creation operators,
the filled state and the vacuum are related by CP, and their electric
charges must be opposite.

The invocation of CP invariance assumes that the theory
we started with is CP-invariant. This means that the symmetric
charge assignment is valid for a vanishing Chern-Simons coupling. But we
know that turning on the Chern-Simons coupling $k$ is equivalent
to shifting the electric charge by $k$ times the
vortex charge. Therefore we conclude that in the
presence of the Chern-Simons coupling the Fock vacuum has electric charge
$$
-\frac{N_f}{2}+k,
$$
while the filled state has charge
$$
\frac{N_f}{2}+k.
$$

Note that because of the parity anomaly, the electric charge is all always integer-valued, whether $N_f$ is even or odd. This a manifestation of
the close relationship between the existence of parity anomaly
and the induced vacuum charge~\cite{NS}.

Now we can analyze the consequences of the Gauss' law constraint.
If all non-zero modes are in their ground state, the constraint
simply says that the total electric charge of the state must be
zero. 
For $k=0$ this implies that a physical state is obtained
by acting with $N_f/2$ zero modes on the vacuum. The number of
such states is
$$
\left(\begin{array}{r} N_f \\ \frac{1}{2} N_f \end{array}\right),
$$
and they transform as an anti-symmetric tensor of $SU(N_f)$
with $N_f/2$ indices. Note that cancellation of global anomalies
requires $N_f$ to be even when $k=0$, so this result makes sense.
For $k$ between
$-N_f/2$ and $N_f/2$ the physical states are obtained by acting
with $N_f/2-k$ zero modes on the vacuum. The corresponding states
transform as an anti-symmetric tensor of $SU(N_f)$ with
$N_f/2-k$ indices. Again global anomaly cancellation ensures
that $N_f/2-k$ is an integer. For $|k|>\frac{N_f}{2}$ there 
are no gauge-invariant states with unit vortex charge and all
non-zero modes in their ground state.

If one does not assume that positive-energy modes are in their
ground state, then one can construct many other states which
satisfy the Gauss' law and have unit vortex charge. However,
such states will have higher energy than the ones discussed
above. 

Now let us consider the more complicated case of $n=2$. For 
simplicity we will set the Chern-Simons coupling to zero and take $N_f$ to
be even.
In the case $n=2$ each fermion has two zero modes which transform as 
a spin-$\frac{1}{2}$ representation of $SU(2)_{rot}$. 
Reasoning based on CP-invariance
tell us that the Fock vacuum has
electric charge $-N_f$. Physical states must
have zero electric charge and are obtained by acting with $N_f$
zero modes (out of a total number of $2N_f$) on the vacuum.
But physical states must also be annihilated by the electric charge density
operator. This is not automatic anymore, because the fermionic zero modes 
are not rotationally invariant. A short computation shows that the
electric charge density operator for the zero modes $\rho_0(x)$ has a piece which transforms as a singlet of $SU(2)_{rot}$ and a piece which transforms 
as a triplet of $SU(2)_{rot}$. The former is simply the average of 
$\rho_0(x)$ over the sphere and is proportional to the total electric charge.
The spin-triplet piece of $\rho_0(x)$ is proportional to the total spin, simply because this is the only spin-triplet one can make out of two 
spin-$1/2$ fermions. Thus the Gauss' law constraint is equivalent to the requirement that the total electric charge as well as the total spin be zero.

For example, for
$N_f=2$, there are six states with zero total electric charge, which
are obtained by acting on the Fock vacuum with two zero modes out of the
available four. Three of these states transform as a vector of $SU(2)_{rot}$
and as singlets of the flavor group $SU(2)_f$ and do not
satisfy the Gauss' law constraint. The remaining three transform
as singlets of $SU(2)_{rot}$ and as a triplet of $SU(2)_f$.
These three states are gauge-invariant. Note that in this case gauge-invariant states transform as an irreducible representation of the flavor group.
For $N_f>2$ this is no longer true, as one can easily check.

\subsection{Quantum numbers of the monopole operators}

In this section we determine the quantum numbers of the simplest
monopole operators, the ones with the lowest conformal dimension for a
given vortex charge.
On general grounds, such an operator lives in a lowest-weight representation
of the conformal group, and its conformal dimension is defined as the
conformal dimension of the lowest-weight vector, or, if we pass to the
radially quantized picture, as the energy of the corresponding state.

Let us begin with the case $n=1$.
As explained above, gauge-invariant states with lowest energy are obtained 
by putting all non-zero modes in their ground states and acting by $N_f/2-k$ 
zero mode creation operators on the vacuum. Obviously such states
transform as an anti-symmetric representation of $SU(N_f)$ with
$N_f/2-k$ indices. It is interesting to note that the usual 
gauge-invariant operators which are polynomials in the fundamental
fields transform trivially under the center of $SU(N_f)$. Indeed,
free fermions have flavor symmetry group $U(N_f)$, and since
we are gauging its $U(1)$ subgroup, the flavor symmetry of QED appears to be
$U(N_f)/U(1)=PU(N_f)=SU(N_f)/\ZZ_{N_f}$. But monopole operators
transform non-trivially under $\ZZ_{N_f}$ (except for $k=\pm N_f/2$).
A very similar effect occurs in $N=2$ $d=4$ super-QCD, where all perturbative
states transform as tensor representations of the flavor group $SO(2N_f)$,
while magnetically charge states transform as spinors~\cite{SW2}.

Other quantum numbers of interest are spin and conformal dimension.
Since the Fock vacuum and the zero modes are rotationally invariant,
the spin of our monopole operator is zero. The dimension is proportional
to the energy of the state. As usual, the definition of the energy
is plagued by ordering ambiguities. However, we have a simple cure:
we can normalize the energy by requiring that the unit operator
have zero dimension. This means that the energy of the ground state
on ${\bf S}^2$ with zero magnetic flux is defined to be zero. 
The energy of any other state can be defined by introducing a UV
regulator, subtracting the regularized energy of the state corresponding
to the unit operator, and then removing the regulator. This procedure gives
a finite answer, which is not sensitive to the precise choice of the
regulator, provided the regulator preserves the symmetries of the
problem. 

In order to make precise the relation between the Casimir energy and
the dimension, recall that the OPE of a spin-zero primary field and
the stress-energy tensor reads:
$$
T_{\mu\nu}(x) O(y) \sim \frac{h}{8\pi} \left(
\frac{\partial}{\partial x^\mu}\frac{\partial}{\partial x^\nu}
\frac{1}{|x-y|}\right) O(y) + \ldots,
$$
where $h$ is the conformal dimension. 
If the stress-energy tensor of free fermions is defined by
$$
T_{\mu\nu}=-\frac{i}{4}\bpsi\left(\gamma_\mu\cD_\nu+
\gamma_\nu\cD_\mu\right)\psi+\frac{i}{4}\left(\cD_\nu\bpsi\gamma_\mu+
\cD_\mu \bpsi\gamma_\nu\right)\psi-g_{\mu\nu}\cL,
$$
then $h_\psi=h_{\bpsi}=1$, the standard normalization.
This implies that in the radially-quantized picture the expectation
value of the stress-energy tensor in the state $\vert O\rangle$
is given by
$$
\langle T_{\mu\nu} dx^\mu\otimes dx^\nu\rangle_O=\frac{h}{4\pi}\left(d\tau^2-
\frac{1}{2}(d\theta^2+\sin\theta^2 d\phi^2)\right).
$$
Thus $h$ is simply the energy of $\vert O\rangle$ with respect to the
Killing vector $\frac{\partial}{\partial \tau}.$ In our case, this means
that the conformal dimension of the monopole operator is the Casimir
energy of $N_f$ free fermions on ${\bf S}^2$ with a magnetic flux.
This Casimir energy for any $n$ is computed in the Appendix. 
For $n=1$ the result is
$$
h_1=N_f \cdot 0.265\ldots.
$$ 
By charge-conjugation symmetry, monopole operator with $n=-1$ has the
same conformal dimension and spin and transforms in the conjugate
representation of the flavor group $SU(N_f).$ 

It is easy to extend the discussion to $n=\pm 2$. As explained in the
previous section, Gauss' law constraint is equivalent to the requirement
of zero spin and zero electric charge. The states with zero
electric charge are obtained by acting with $N_f$ zero modes (out of total
number of $2N_f$ zero modes) on the Fock vacuum. These states transform
as an anti-symmetric tensor of $SU(2N_f)$ with $N_f$ indices.
Gauge-invariant states are obtained by decomposing this representation
with respect to the $SU(2)_{rot}\times SU(N_f)$ subgroup and separating
out $SU(2)_{rot}$-singlets. In general, gauge-invariant states transform
as a reducible representation of $SU(N_f)$. One can easily show that
the dimension of this reducible representation is 
$$
\left(\begin{array}{c} \frac{1}{2}N_f^2+N_f-1 \\ \frac{1}{2} N_f \end{array}\right)
$$
The conformal dimension
of the corresponding monopole operators is the Casimir energy of
free fermions in a background magnetic field. Numerically, it is
given by
$$
h_2=N_f\cdot 0.673\ldots.
$$
It is interesting to note that $2h_1<h_2$ (at least for large $N_f$).
Therefore the
OPE of two monopole operators with $n=1$ and the lowest conformal dimension contains only terms with positive powers of $|x_1-x_2|$.

\section{Discussion}

In this paper we have constructed local operators in an interacting
3d CFT which carry vortex charge and therefore create 
Abrikosov-Nielsen-Olesen vortices in the Higgs phase. We have shown that 
for large $N_f$ such operators have conformal dimensions of order $N_f$. For the case of unit vortex charge, we showed that the operator with the
lowest possible dimension has zero spin and transforms in a non-trivial
representation of the flavor group. An important tool in this analysis
is $1/N_f$ expansion. 

The idea that vortex-creation operators can be studied in the large $N_f$
limit has been previously proposed in Ref.~\cite{KS}. The approach taken there
was to integrate out the matter fields, and then perform a duality
transformation on the effective action for the gauge field. Then the vortex-creation operator is defined as the exponential of the dual photon. 
One drawback of this approach is that it is easy to miss fermionic
zero modes, and consequently to misidentify the quantum numbers of the
vortex-creating operator. It is preferable to keep the matter fields,
and to identify a vortex-creating operator by the property that its insertion
causes a change in the topology of the gauge field. As we have seen above,
this definition can be made concrete by using radial quantization and
large $N_f$ expansion.

Our main motivation for studying vortex-creation operators was the hope
that this would enable us to give a constructive proof of 3d mirror
symmetry. It is straightforward to apply the methods of this paper
to 3d gauge theories with $N=2$ and $N=4$ supersymmetry.
The results and their implications for mirror symmetry will be reported
in a forthcoming publication.

It is natural to wonder if our approach to the construction of 
topological disorder operators has an analogue in four dimensions. In three
dimensions, we defined the vortex charge of a local operator as the first
Chern class of the gauge bundle evaluated on an ${\bf S}^2$ surrounding the
insertion point. In four dimensions, we have ${\bf S}^3$ instead of 
${\bf S}^2$, and since characteristic classes of vector bundles are even-dimensional, it appears impossible to define a similar topological charge for local operators. On the other hand, a B-field on an ${\bf S}^3$
can have non-trivial topology, since its field-strength is a 3-form.
Thus, if there were an interacting 4d CFT involving a B-field, one could
define local operators which create topological disorder. In order for this to work, the field-strength 3-form must have dimension $3$, so that its
dual is a conserved primary current. Note that in the theory of a free
B-field, the field-strength has dimension $2$. In this case the dual current, although conserved, is not a primary, but a gradient of a free scalar. 
Thus in order to define a conformally-invariant
topological charge, the 4d CFT {\it must} be interacting. Unfortunately,
no such theory is known at present. Perhaps there exists a duality-symmetric 
reformulation of $N=4$ $d=4$ super-Yang-Mills which involves B-fields,
and in which both W-bosons and dual W-bosons are described by 
topological disorder operators.

After the first version of this paper was posted on the arXive, we learned that topological disorder operators in 3d have
been previously considered by G.~Murthy and S.~Sachdev~\cite{MurthySachdev}. The model considered there was the $\CC\PP^N$ model
in three dimensions. This theory is not renormalizable and requires an ultraviolet cut-off. It has a critical point 
separating the ordered phase, where the sigma-model field has an expectation value, and the disordered phase, where
the correlators decay exponentially. Unlike in 3d QED, there are dynamical topological defects in the $\CC\PP^N$ model
(so-called hedgehogs). But at the critical point and in the limit of large $N$ their density vanishes, and the situation
becomes very similar to that in 3d QED. In particular, the critical exponents computed in Ref.~\cite{MurthySachdev}
can be interpreted as scaling dimensions of hedgehog operators. It is interesting to note that the approach
to computing these scaling dimensions taken by Murthy and Sachdev is rather different from ours. Instead of evaluating
the expectation value of the stress-energy tensor in the presence of topological disorder operators, they in essence
compute the 2-point function of these operators. This is somewhat obscured by the fact that Murthy and Sachdev
map both operator insertions to infinity. As a result, the distance dependence of the 2-point function is traded for
an anomalous dependence of the partition function on the ratio of the ultraviolet cut-off $\Lambda$ and the infrared cut-off
$\Delta$. This method could be used in 3d QED as well. In fact, M.~J.~Strassler and one of the authors of the present
paper (A.K.) have contemplated such a route to computing scaling dimensions of monopole operators in 3d QED and SQED, 
but were discouraged by apparent technical difficulties. It would be interesting to rederive the results of the
present paper using the approach of Ref.~\cite{MurthySachdev}.

\section*{Appendix}

\subsection*{Monopole harmonics}

To solve for the energy spectrum of free fermions on ${\bf S}^2$ with
a magnetic flux, we will use the fact that this system is related by a conformal transformation to the Dirac equation on $\RR^3$ in the monopole background. This allows us to use the machinery of ``monopole harmonics'' developed by Wu and Yang~\cite{WY}. 

The three-dimensional Dirac operator on flat $\RR^3$ is 
given by
\begin{equation*}
iD=-{\vec{\sigma}}\cdot{\vec{\pi}},
\end{equation*}
where $\sigma_x$, $\sigma_y$, and $\sigma_z$ are the Pauli matrices, and 
${\vec{\pi}}=\vec{p}+\vec{A}$, with $\vec{p}$ being the momentum 
operator. Following Ref.~\cite{WY}, let us define the generalized orbital angular momentum operator as
\begin{equation*}
\vec{L}=\vec{r}\times\vec{\pi}-\frac{q\vec{r}}{r}
\end{equation*}
with $q=-eg=n/2$.
It is straightforward to check that $\vec{L}$ defined this way satisfies 
the angular momentum algebra:
\begin{eqnarray}
\left[ L_j, x_k \right]=i\epsilon_{jkm}x_m,\nonumber\\
\left[ L_j, \pi_k \right]=i\epsilon_{jkm}\pi_m,\nonumber\\
\left[ L_j, L_k \right]=i\epsilon_{jkm}L_m.\nonumber
\end{eqnarray}
We define the total angular momentum as
\begin{equation*}
\vec{J}=\vec{L}+\frac{\vec{\sigma}}{2}.
\end{equation*}

We can take $r, {\vec{L}}^2, {\vec{J}}^2,$ and $J_z$ as 
a complete set of observables (it is easy to check that they commute and 
are all self-adjoint with respect to the usual inner product). It can be checked that $\left[ \vec{J}, iD \right]=0$, but $\left[\vec{L}^2, iD\right]\neq0$. 
However, this is good enough, because as we will see later, to find the 
eigenvalues of $iD$ we only need to diagonalize an operator in
a two-dimensional space.

The monopole harmonics $Y_{q,l,m}(\theta, \phi)$ constructed in 
Ref.~\cite{WY} satisfy
\begin{align*}
{\vec{L}}^2Y_{q,l,m}=l(l+1)Y_{q,l,m},\quad L_zY_{q,l,m}=mY_{q,l,m},\\
l=|q|,|q|+1,|q|+2,\ldots,\quad m=-l,\ldots,l.
\end{align*}
The simultaneous eigenfunctions of $\left\{ 
{\vec{L}}^2, {\vec{J}}^2, J_z\right\}$ will be denoted by $\phi_{ljm_j}$
and are given by
\begin{eqnarray}
\phi_{ljm_j}=\begin{pmatrix}
\sqrt{\frac{l+m+1}{2l+1}} Y_{q,l,m} \\
\sqrt{\frac{l-m}{2l+1}} Y_{q,l,m+1}
             \end{pmatrix}
\ {\rm for}\ j=l+\frac{1}{2}\ (m_j=m+\frac{1}{2}),\nonumber\\
\phi_{ljm_j}=\begin{pmatrix}
-\sqrt{\frac{l-m}{2l+1}} Y_{q,l,m}\\
\sqrt{\frac{l+m+1}{2l+1}} Y_{q,l,m+1}
             \end{pmatrix}
\ {\rm for}\ j=l-\frac{1}{2}\ (l\neq 0, m_j=m+\frac{1}{2}).\nonumber
\end{eqnarray}
They satisfy
\begin{eqnarray}
{\vec{L}}^2\phi_{ljm_j}=l(l+1)\phi_{ljm_j},\nonumber\\
{\vec{J}^2}\phi_{ljm_j}=j(j+1)\phi_{ljm_j},\nonumber\\
J_z\phi_{ljm_j}=m_j\phi_{ljm_j}.\nonumber
\end{eqnarray}
We can summarize the possible value of $l, j, m_j$ as follows:
\begin{align*}
&\bullet\ j=|q|-\frac{1}{2},\ |q|+\frac{1}{2},\ |q|+\frac{3}{2},\  
|q|+\frac{5}{2},\ldots\\
&\ ({\rm for}\ q=0,\ j=|q|-\frac{1}{2} {\rm is\ not\ allowed});\\
&\bullet\ {\rm if}\ j=|q|-\frac{1}{2},\ {\rm then}\ l=j+\frac{1}{2}=|q|,
\ {\rm otherwise}\ l=j\pm \frac{1}{2};\\
&\bullet\ m_j= -j, -(j-1),\ldots, j-1, j.
\end{align*}
Any wave-function can be expanded as
\begin{equation*}
\psi(\vec{r})=\sum_{l,j,m_j}R_{ljm_j}(r)\phi_{ljm_j}(\theta,\phi).
\end{equation*}
Note that while $\phi_{ljm_j}$ is a two-component spinor, $R_{ljm_j}$ is 
just a scalar.

Now we may write $iD$ in terms of the angular momenta.
Define $\sigma_r$ as
\begin{equation*}
\sigma_r=\frac{\vec{\sigma}\cdot\vec{r}}{r}.
\end{equation*}
One can show that 
\begin{equation*}
\sigma_r (iD)=i\frac{\partial}{\partial 
r}-i\frac{1}{r}\vec{\sigma}\cdot\vec{L}-iq\frac{\sigma_r}{r},
\end{equation*}
where we made use of the fact that
\begin{equation*}
\left(\vec{\sigma}\cdot\vec{G}\right)\left(\vec{\sigma}\cdot\vec{K}\right) 
=\vec{G}\cdot\vec{K}+i\vec{\sigma}\cdot\left(\vec{G}\times\vec{K}\right)
\end{equation*}
for any $\vec{G}$ and $\vec{K}$ that commute with $\vec{\sigma}$.
Now using the fact that ${\sigma_r}^2=1$, we have
\begin{eqnarray}
iD&=&\sigma_r\sigma_r(iD)=i\sigma_r\frac{\partial}{\partial 
r}-i\frac{\sigma_r}{r}\vec{\sigma}\cdot\vec{L}-iq\frac{1}{r}\nonumber\\
&=&i\sigma_r\frac{\partial}{\partial r}-i\frac{\sigma_r}{r}({\vec{J}}^2 
-{\vec{L}}^2-\frac{3}{4})-iq\frac{1}{r}.\nonumber
\end{eqnarray}
Thus the Dirac Lagrangian on $\RR^3$ in the presence of a monopole
can be written as
$$
{\cal L}_{R^3}=\frac{i}{r}
{\bar{\psi}}\sigma_r\left(r\frac{\partial}{\partial r}
-(\vec{J}^2-\vec{L}^2-\frac34)-q\sigma_r\right)\psi. 
$$
Setting $r=e^\tau$ and performing a Weyl rescaling 
$$
g_{\mu\nu}\ra e^{-2\tau} g_{\mu\nu},\quad \psi,\bar{\psi}\to e^{-\tau}\psi, e^{-\tau}\bar{\psi},\quad \vec{A}\to\vec{A},
$$
we obtain the Lagrangian on ${\bf S}^2\times \RR$:
$$
{\cal L}_{{\bf S}^2\times \RR}=i\bar{\psi}\sigma_r\left(
\frac{\partial}{\partial\tau}-
(\vec{J}^2-\vec{L}^2+\frac14)-q\sigma_r\right)\psi.
$$
Note that the norm
$$
\int_{{\bf S}^2}\phantom{}r^2\, d\Omega\,\bar{\psi}\sigma_r\psi
$$
on $\RR^3$ is transformed to the norm
$$
\int_{{\bf S}^2}\phantom{}\bar{\psi}\sigma_r\psi
$$
on ${\bf S}^2\times \RR.$

Taking into account the above results, the Euclidean
equation of motion for $\psi$ is 
\begin{eqnarray}
&\ &
\frac{dR_{ljm_j}(\tau)}{d\tau}-\left(j(j+1)-l(l+1)
+\frac{1}{4}\right)R_{ljm_j}(\tau) \nonumber\\
&\ &- \sum_{l'j' m'_j}q R_{l'j'm'_j}(\tau)
\langle ljm_j\vert\sigma_r\vert l'j'm'_j\rangle  =0,\nonumber
\end{eqnarray}
where we have used $\langle ljm_j\vert\sigma_r\vert l'j'm_j'\rangle$ 
to denote $\int d\Omega \phi_{ljm_j}^\dagger\sigma_r\phi_{l'j'm_j'}.$

Now the identity $\left[\vec{J}, \sigma_r\right]=0$ tells us that
\begin{equation*}
\langle ljm_j\vert\sigma_r\vert l'j'm_j'\rangle= 
\delta_{jj'}\delta_{m_jm_j'}\langle ljm_j\vert\sigma_r\vert l'jm_j\rangle,
\end{equation*}
and thus the eigenvalue equation becomes, for any given $j,m_j,$
\begin{eqnarray}
&\ &
\frac{dR_{ljm_j}(\tau)}{d\tau}-\left(j(j+1)-l(l+1)
+\frac{1}{4}\right)R_{ljm_j}(\tau) \nonumber\\
&\ &- \sum_{l'}q R_{l'jm_j}(\tau)
\langle ljm_j|\sigma_r|l'jm_j\rangle  =0.\nonumber
\end{eqnarray}
Let us suppress the $j,m_j$ indices, and denote 
$R_{(l=j-\frac{1}{2})jm_j}$ by $R^a$, 
$\vert l=j-\frac{1}{2},jm_j\rangle$ by $\vert a\rangle$, $R_{(l=j+\frac{1}{2})jm_j}$ by $R^b$, $\vert l=j+\frac{1}{2},jm_j\rangle$  
by $\vert b\rangle$, $\langle a\vert \sigma_r\vert a\rangle$ by 
$\sigma_{aa}$, $\langle a\vert\sigma_r\vert b\rangle$ by $\sigma_{ab}$,
$\langle b\vert\sigma_r\vert a\rangle$ by 
$\sigma_{ba}$, and $\langle b\vert\sigma_r\vert b\rangle$ by $\sigma_{bb}$. Then for any given 
$j,m_j$, we have two coupled first-order differential equations:
$$
\frac{dR^a(\tau)}{d\tau}=\left(j+\frac{1}{2}\right)R^a(\tau)
+q(\sigma_{aa}R^a(\tau)+\sigma_{ab}R^b(\tau)),
$$
$$
\frac{dR^b(\tau)}{d\tau}=-\left(j+\frac{1}{2}\right)R^b(\tau)
+q(\sigma_{bb}R^b(\tau)+\sigma_{ba}R^a(\tau)).
$$

A straightforward calculation of the matrix elements $\sigma_{aa}$, $\sigma_{ab}$, and $\sigma_{bb}$ gives 
\begin{equation*}
\sigma_{aa}=\frac{-q}{j+\frac{1}{2}},\ 
\sigma_{bb}=\frac{q}{j+\frac{1}{2}},\ 
\sigma_{ab}=-\sqrt{1-\left(\frac{q}{j+\frac{1}{2}} \right)^2}, 
\end{equation*}
and of course $\sigma_{ba}=\sigma_{ab}^*=\sigma_{ab}$.

\subsection*{Energy spectrum and Casimir energy for fermions on ${\bf S}^2$}

The energy spectrum can be read off from the behavior of the solutions
as a function of $\tau$: a solution with energy $E$ behaves as $e^{-E\tau}.$
The results are as follows.

\noindent {\it Case (i): $q=0.$}

The two equations decouple, and we find 
$$
R^a(\tau)=C^ae^{(j+\frac{1}{2})\tau},\quad R^b(\tau)=
C^be^{-(j+\frac{1}{2})\tau},
$$
where $C^a$ and  $C^b$ are integration constants, and 
$j=\frac{1}{2},\frac{3}{2},\frac{5}{2}\dots$ 
There are no zero-energy solutions.

\noindent {\it Case (ii): $q\neq 0$, $j=|q|-\frac{1}{2}.$}

In this case, there is no such thing as $R^a$ (because $l$ cannot be 
$j-\frac{1}{2}=|q|-1$). So the first equation is absent, and the 
second equation gives:
$$
R^b(\tau)=C,
$$
with an arbitrary constant $C$. This solution has zero energy 
and degeneracy $2j+1=2|q|$.

\noindent {\it Case (iii): $q\neq 0$ and $j=|q|-\frac12+p> |q|-\frac{1}{2}$,
$p=1,2,\dots .$}

In this case, the two equations are coupled but easy to solve by 
eliminating one of the two unknowns. The result is
\begin{eqnarray}
R^a(\tau)&=&q C_1e^{\tau\sqrt{\left(j+\frac{1}{2}\right)^2-q^2}} 
+q C_2e^{-\tau\sqrt{\left(j+\frac{1}{2}\right)^2-q^2}},\nonumber\\
R^b(\tau)&=&\left[\sqrt{\left(j+\frac{1}{2}\right)^2-q^2}
-\left(j+\frac{1}{2}\right)\right]
C_1e^{\tau\sqrt{\left(j+\frac{1}{2}\right)^2-q^2}}\nonumber\\
&&+\left[\sqrt{\left(j+\frac{1}{2}\right)^2-q^2}+\left(j+\frac{1}{2}\right)
\right]C_2e^{-\tau\sqrt{\left(j+\frac{1}{2}\right)^2-q^2}},\nonumber
\end{eqnarray}
where $C_1$ and $C_2$ are integration constants.
The corresponding energies are 
$$
E_p=\pm\sqrt{\left(j+\frac12\right)^2-q^2}=\pm\sqrt{2|q|p+p^2},
$$
with degeneracies $2j+1=2|q|+2p$. Note that the spectrum is symmetric
under $q\to -q$.

The regularized Casimir energy is given by
$$
E_{reg}(\beta)=-\sum_{p=0}^\infty (2p+|n|)\sqrt{p^2+p|n|}
e^{-\beta\sqrt{p^2+p|n|}}.
$$
We renormalize it by requiring that the Casimir energy of the vacuum with
$n=0$ be zero. That is, we subtract from the above sum a similar sum
with $n=0$, and then take the limit $\beta\ra 0$.
Using the Abel-Plana summation formula
$$
\sum_{p=0}^\infty F(p)=\frac12 F(0)+\int^\infty_0 dx\, F(x)+
i\int^\infty_0 dt\, \frac{F(it)-F(-it)}{e^{2\pi t}-1},
$$
we obtain a finite answer for the Casimir energy:
\begin{multline}
E_{Casimir}=\frac16 \sqrt{1+|n|}(|n|-2)
+\\
4\,\textrm{Im}\int^\infty_0 dt\phantom{}
\left[ \left(it+\frac{|n|}{2}+1\right)\sqrt{\left(it+\frac{|n|}{2}+1\right)^2-
\frac{n^2}{4}}\right]\frac{1}{e^{2\pi t}-1}.\nonumber
\end{multline}
Here one needs to take the branch of the square root which is positive on the
positive real axis. The integral cannot be expressed in terms of elementary functions, but can be easily evaluated numerically for any $n$.

\section*{Acknowledgments}
This work grew out of attempts by one of the authors (A.K.) and 
M.~J.~Strassler to improve on the last section of Ref.~\cite{KS}. A.K.
would like to thank M.~J.~Strassler for numerous discussions which
helped to realize the importance of fermionic
zero modes. We also would like to thank J.~Maldacena, T.~Okuda, and
H.~Ooguri for useful conversations, and S.~Sachdev for informing us about Ref.~\cite{MurthySachdev}.
This work was supported in part by a 
DOE grant DE-FG03-92-ER40701.

\end{document}